\documentstyle[twocolumn,aps,epsf]{revtex}

\begin{document}
\draft
\title{Measurement of the Zero Crossing  in a
Feshbach Resonance of Fermionic $^6$Li}
\author{K. M. O'Hara, S. L. Hemmer, S. R. Granade, M. E. Gehm,
and J. E. Thomas}
\address{Physics Department, Duke University,
Durham, North Carolina 27708-0305}
\author{V.Venturi, E. Tiesinga, and C.J. Williams }
\address{Atomic Physics Division, National Institute of Standards and Technology,
Gaithersburg, Maryland 20899-8423}
\date{\today}
\wideabs{\maketitle
\begin{abstract}
We measure a zero crossing in the scattering length of a mixture
of the two lowest hyperfine states of $^6$Li. To locate the zero
crossing, we monitor the decrease in temperature and atom number
arising from evaporation in a CO$_2$ laser trap as a function of
magnetic field $B$. The temperature decrease and atom loss are
minimized for $B=528\pm 4$ G (1 G=$10^{-4}$ T), consistent with no
evaporation. We also present preliminary calculations using
potentials that have been constrained by the measured zero
crossing and locate a broad Feshbach resonance at approximately
860 G, in agreement with previous theoretical predictions. In
addition, our theoretical model predicts a second and much
narrower Feshbach resonance near 550 G.
\end{abstract}
\vspace*{.125in} }

Recently, three groups have predicted the possibility of
high-temperature superfluid transitions arising from the strong
pairing of a two-component Fermi gas in the vicinity of a Feshbach
resonance~\cite{Holland,Timmermans,Griffin}. Transition
temperatures  as large as half the Fermi temperature are
expected~\cite{Holland,Timmermans,Griffin,Kokkelmans}. This is a
much larger fraction than in standard Bardeen-Cooper-Schriefer
(BCS) theory. One promising atomic system is $^{40}$K, for which a
Feshbach resonance has been observed~\cite{JinFeshbach} and
degeneracy has been achieved~\cite{Jin,Roati}. Another promising
candidate is $^6$Li, which is predicted to exhibit magnetically
tunable Feshbach resonances in mixtures of the two lowest
hyperfine states. Such transition temperatures are within the
reach of current experiments involving degenerate $^6$Li Fermi
gases~\cite{Hulet,Salomon,Granade,Hadzibabic}. Hence, locating the
resonance is of importance for studies of atomic gas analogs of
high temperature superconductivity.

As is well known, a Feshbach resonance arises in a colliding
two-atom system when the energy of the incoming open elastic
channel is magnetically tuned into resonance with a bound
molecular state of an energetically closed channel. The tuning
dependence arises from the difference in the magnetic moments of
the open and closed channels~\cite{Verhaar}. In the vicinity of a
Feshbach resonance, the $s$-wave scattering length $a(B)$ is
described approximately by~\cite{Kokkelmans}
\begin{equation}
a(B)=a_{bg}\left(1-\frac{\Delta B}{B-B_0}\right) ,
\label{eq:Feshbach}
\end{equation}
where $B$ is the applied field, $a_{bg}$ is the background
scattering length, $B_0$ is the field at which the resonance
occurs, and $\Delta B$ is proportional to the strength of the
coupling between the open and closed channels.

For an ultracold gas which is a mixture of the two lowest
hyperfine states, $|1\rangle$ and $|2\rangle$, of $^6$Li (the
$|1/2,\pm 1/2\rangle$ states in the low-field, total angular
momentum $|f,M_f\rangle$ basis),  $s$-wave collisions only occur
between atoms in these two different spin states. Hence, at
sufficiently low temperatures, the scattering length of the
mixture is the scattering length of this collision. The incoming
channel of the Feshbach resonance in this mixture is predominantly
triplet in character~\cite{Elastic}. Thus near the resonance,
$a_{bg}$ is nearly equal to the triplet $a \, ^3\Sigma_u^+$
scattering length which is large and negative~\cite{s-wave}.  The
vibrational level that gives rise to the resonance is due to the
most weakly-bound vibrational level ($v=38$) of the singlet $X \,
^1\Sigma_g^+$ potential. The zero field binding energy of the
$v=38$  vibrational level has not been directly measured, and the
uncertainty in the binding energy can lead to significant
theoretical uncertainty in the location of the Feshbach resonance
and the associated zero crossing.
\begin{figure}
\begin{center}\
\epsfysize=55mm \epsfbox{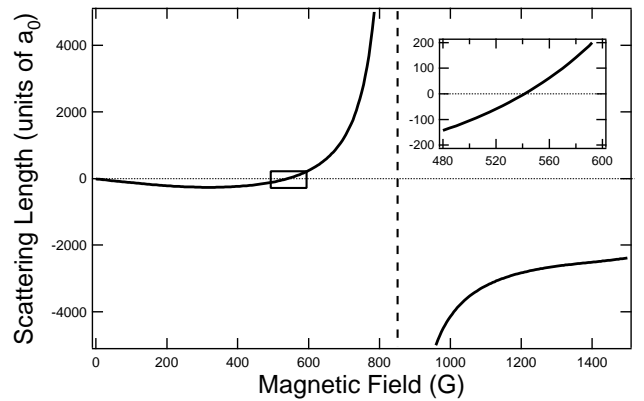}
\end{center}
 \caption{Analytic model
 of the scattering length for a mixture of the two lowest
 hyperfine states of $^6$Li as a function of magnetic
 field showing the zero crossing (inset). ($1 \, a_0=0.0529177$ nm) } \label{fig:1}
\end{figure}

Figure~\ref{fig:1} shows an analytic model~\cite{Kokkelmans2} of
the scattering length as a function of magnetic field between 0 G
and 1500 G, and is based on adjusting model parameters to fit the
results of previous multichannel
calculations~\cite{Kokkelmans,Elastic}. The scattering length is
predicted to cross zero between 500 G and 600 G for a mixture of
states $|1\rangle$ and $|2\rangle$ in $^6$Li and in
Fig.~\ref{fig:1} is shown to cross the horizontal axis near 540 G.

Previously, we have shown~\cite{Granade,O'Hara1} that the
scattering length for a mixture of the $|1\rangle$ and $|2\rangle$
states is very small at zero magnetic field in agreement with
predictions~\cite{Elastic}. This was accomplished by confining the
mixture in an ultrastable CO$_2$ laser trap for several hundred
seconds at $\simeq 100\,\mu$K with negligible evaporation. At
magnetic fields of 100-300 G, we have observed large evaporation
rates suitable for achieving degeneracy~\cite{Granade}. Further,
we observed scattering cross sections corresponding to a
scattering length with a magnitude of 540 $a_0$ at a magnetic
field of 8 G in a mixture of the $|1\rangle$ and
$|3\rangle=|f=3/2,M_f=-3/2\rangle$
 hyperfine states. The measured
scattering cross section is consistent with a calculation based on
the predicted~\cite{s-wave} singlet (45.5 $a_0$) and triplet
(-2160 $a_0$) scattering lengths~\cite{O'Hara2}.

In this paper, we  observe the vanishing of the elastic scattering
cross section, and hence the vanishing of the elastic scattering
length, for a mixture of the  $|1\rangle$ and $|2\rangle$
hyperfine states of $^6$Li in a magnetic field. This provides a
first step in determining the location of a Feshbach resonance in
this system. To determine the location of the zero crossing, we
monitor the temperature drop and atom loss after 40 seconds of
free evaporation  in a stable CO$_2$ laser trap as a function of
applied magnetic field.

Our $^6$Li experiments employ a stable CO$_2$ laser trap with a
single focused and retroreflected beam~\cite{Granade}. The atoms
evaporate from the trap, which has a fixed depth of $\simeq 0.75$
mK and a geometric mean trap frequency of $\simeq 2.5$ kHz . Since
the scattering length for the $^6$Li $|1\rangle$ and $|2\rangle$
mixture is extremely small at zero magnetic field, evaporation is
turned on and off simply by applying or not applying a magnetic
field.

The CO$_2$ laser trap is continuously loaded from a $^6$Li
magneto-optical trap (MOT)~\cite{Granade}. Typically, $2\times
10^6$ atoms in a nominally 50-50 mixture of the $|1\rangle$ and
$|2\rangle$ states are initially contained in the trap at a
temperature of 140 $\mu$K, as determined by time-of-flight
absorption imaging. Holding the atoms at zero magnetic field for
an additional 40 seconds does not produce a measurable temperature
change in the time-of-flight images, consistent with the
prediction of a small scattering length at $B=0$ G. In  free
evaporation,  the number of atoms decreases rapidly to
approximately 1/3 of the initial value as the hottest atoms are
lost~\cite{Granade}.

To measure the magnetic field dependence of the  evaporation rate,
a pair of high field magnets with currents of 0-240 A are used to
generate a uniform field of 0-1100 G in the trap region. The
magnets have a diameter of $\simeq 20$ cm and are located $\simeq
12.5$ cm from the trap, producing negligible field curvature.
 The MOT gradient field is generated by reversing  the current
in one magnet for the trap loading stage. Each of the magnets is
powered by a pair of current-regulated supplies
which dissipate up to 10 kW per magnet.

\begin{figure}
\begin{center}\
\epsfysize=60mm \epsfbox{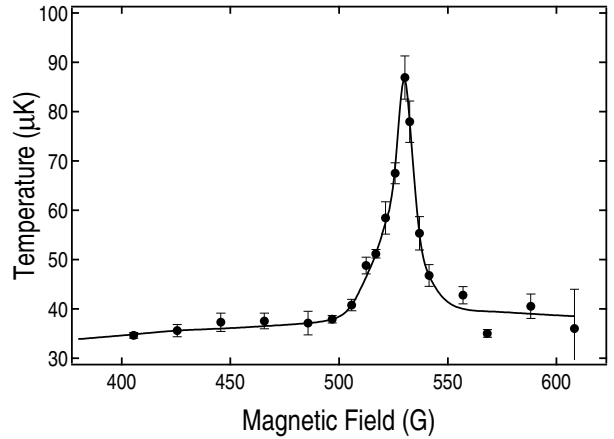}
\end{center}
 \caption{Temperature after 40 seconds of free evaporation as a
 function of magnetic field. The maximum temperature occurs
 near the zero crossing  of the scattering length where the
 evaporation rate vanishes. A curve has been
 added to guide the eye.} \label{fig:2}
\end{figure}

In the experiments, the trap is initially  loaded at zero magnetic
field. Then a uniform magnetic field is applied for 40 seconds,
during which time the atoms evaporate from the trap. Finally, the
field is returned to 0 G and the temperature and number are
measured by time-of-flight absorption imaging~\cite{Granade}. The
final temperature obtained as a function of magnetic field between
400 G and 610 G is shown in Fig.~\ref{fig:2}.

For magnetic fields far from the peak at 530 G, the final
temperature of the atoms is 35 $\mu$K, well below the Doppler
cooling limit of 140 $\mu$K observed  when the magnetic field is
held at 0 G for 40 seconds. As the magnetic field approaches the
peak, the observed temperature increases to 87 $\mu$K, which is
below the loading temperature, but well above the minimum
temperature observed far from the zero crossing. The drop in the
maximum temperature from 140 $\mu$K to 87 $\mu$K arises from the
0.3 seconds during which the current-regulated supplies change
current. Since the field is switched  from 0-530 G and back to 0
G, the atoms evaporate for an extra period of  0.6 seconds. By
switching the magnetic field from 0-530 G and back to 0 G without
the 40 second holding time, we find that the final temperature is
90 $\mu$K. The remaining 3 $\mu$K temperature drop over the 40
second holding period at 530 G can be accounted for by assuming a
small error in determining the zero crossing.

\begin{figure}
\begin{center}\
\epsfysize=60mm \epsfbox{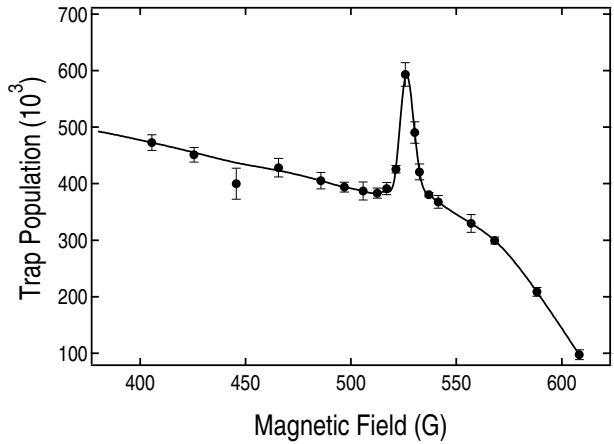}
\end{center}
 \caption{Number of atoms remaining after 40 seconds of free evaporation
 as a function of magnetic
 field. The maximum number occurs near the zero crossing
 of the scattering length where the evaporation rate vanishes. A curve has been
 added to guide the eye.} \label{fig:3}
 \end{figure}

To confirm  that the peak in temperature arises from a
reduced cooling rate and not from heating, we also extracted the
number of remaining atoms from the time-of-flight data,
Fig.~\ref{fig:3}. Consistent with a reduced evaporation rate, we
observe a peak in the number at 526 G, nearly overlapping the peak
in the temperature. Further, near the zero crossing field, we
observe a $\simeq 30$\% increase in the number of atoms with a
negligible temperature change when the holding time is decreased
from 40 to 20 seconds, indicating a negligible heating rate.

We also observe a decrease in the remaining trap population as the
magnetic field is tuned above 600 G, as shown in Fig.~\ref{fig:3}.
This appears to arise from a broad magnetic field dependent loss
feature which we observe to be near 650 G. In this region, we
observe loss and heating over time scales of a second in a sample
at a temperature of $5\,\mu$K and a density of $3\times
10^{13}/{\rm cm}^3$. This loss feature also has been observed by
Dieckmann et al.~\cite{Dieckmann}.  Note that inelastic loss rates
should not cause heating in the region of the zero crossing, since
the elastic scattering cross section is small and secondary
scattering does not occur.  However, the rapid increase in loss
with increasing magnetic field may be partially responsible for
the 4 G shift between the peaks in the number (Fig.~\ref{fig:3})
and temperature (Fig.~\ref{fig:2}).

The magnetic field is calibrated  using optical absorption
resonances to measure the splitting between the $M_J=-3/2$ and
$M_J=-1/2$ levels of the $2P_{3/2}$ excited electronic state.
Acousto-optic (A/O) modulators are used to produce two
copropagating probe beams differing by a tunable frequency up to
$\simeq 1000$ MHz. One beam is {\bf z} polarized and interacts
with the $2S_{1/2}$, $M_J=-1/2\rightarrow$ $2P_{3/2}$, $M_J=-1/2$
transition originating from the $|2\rangle$ ground hyperfine
state. The other beam is {\bf x} polarized and interacts with the
$2S_{1/2}$, $M_J=-1/2\rightarrow$ $2P_{3/2}$, $M_J=-3/2$
transition originating from the $|2\rangle$ ground hyperfine
state. For fields near the zero crossing, note that
$|2\rangle\simeq |m_s=-1/2,m_I=0\rangle$.

 At the magnetic field corresponding to
the zero crossing, the {\bf z} polarized beam is blocked and the
dye laser frequency is adjusted to tune the  {\bf x} polarized
beam into resonance. The resonance peak is determined by spatially
integrating a time-of-flight absorption image for each dye laser
frequency.  This method avoids degradation of the image by
diffraction arising from the small radial profile of the trapped
atoms. The integrated absorption signal is insensitive to phase
shifts arising from a nonzero index of refraction when the laser
is detuned.  The laser is then locked at the resonance frequency.
Then the {\bf x} polarized beam is blocked and a similar method is
used to tune the {\bf z} polarized beam into its resonance  by
adjusting an A/O frequency. The excited state frequency splitting
is measured from the beat frequency of the two beams  using a
diode detector. At the magnetic field for the atom loss minimum,
Fig.~\ref{fig:3}, we find a beat frequency of 994 MHz.

The magnetic field corresponding to the measured beat frequency is
determined by calculating the expected splitting versus
magnetic field, including the Zeeman, hyperfine, and fine
structure mixing contributions. From the 994 MHz splitting, the
atom number peak (loss minimum) is determined to be at 526 G. The
minimum temperature decrease is then determined to be at 530 G.
Note that at 526 G, magnetic field-induced fine structure mixing
causes a quadratic Zeeman shift of $\simeq 12$ MHz in Li, due to
the small fine-structure splitting.

The uncertainty in the field determination arises primarily from
the widths of the observed optical absorption resonances, which
are about 10 MHz or 5 G in the vicinity of the zero crossing. The
shape of each optical resonance exhibits some saturation arising
from optical pumping. However, the centers are easily located with
a statistical uncertainty of less than $\pm 2$ G. We report the
location of the zero crossing as 528 G, the average of the
locations of the peaks in the temperature, Fig.~\ref{fig:2}, and
number, Fig.~\ref{fig:3}. The uncertainty, $\pm 4$ G, is the root
mean square or combined standard uncertainty of the systematic and
statistical uncertainties. This value is dominated by the
systematic  error which we estimate as the shift between the
temperature and number peaks (4 G).

The measurement of the zero crossing can be used to further
constrain the singlet potential and hence the singlet scattering
length. Therefore we present preliminary results of a
coupled-channel calculation that attempts to optimize the singlet
potential. The interaction potentials in these calculations are
constructed using the short-range singlet potential of
Ref.~\cite{Cote:1994}, the short-range triplet potential of
Ref.~\cite{Linton:1999}, the long-range dispersion coefficients of
Ref.~\cite{Yan:1996}, and the methodology of Ref.~\cite{Cote:1994}
to connect the short- and long-range potentials. In addition, the
triplet potential has been modified to be consistent with the
measured binding-energy of the most weakly bound state of the $a
\, ^3\Sigma_u^+$ state of $^6$Li~\cite{s-wave}. The singlet
potential has been made to agree with the most recent prediction
of the $^6$Li singlet scattering
length~\cite{s-wave,vanAbeelen:1997}. Subtle mass-polarization
corrections which affect the dissociation energy for ${}^6$Li and
${}^7$Li have been neglected.

\begin{figure}
\begin{center}\
\epsfysize=60mm \epsfbox{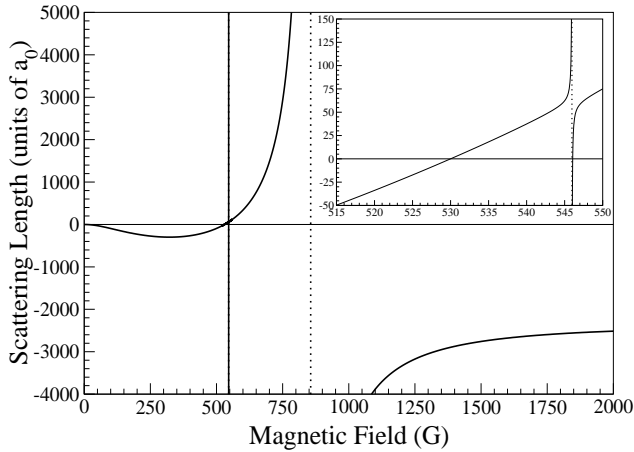}
\end{center}
 \caption{Scattering length for collisions of $^6$Li atoms in the two
 lowest hyperfine states as a function of magnetic field. The
 singlet potential has been fixed to obtain a zero scattering length at
 530 G. \label{fig:4}}
 \end{figure}

The scattering length of the singlet potential can be varied
within the uncertainty given by
Ref.~\cite{s-wave,vanAbeelen:1997}. The corresponding uncertainty
of the magnetic field position of the zero in the scattering
length does not exclude our observed zero crossing. In
Fig.~\ref{fig:4} we show the scattering length as a function of
magnetic field, where the scattering length of the singlet
potential has been fixed to obtain a zero crossing at 530 G. Close
examination of the zero crossing region has also revealed a narrow
$s$-wave Feshbach resonance not found in previous calculations.
This resonance is positioned at approximately 550G between the
zero crossing and the peak of the broad resonance which is
centered at approximately 860G. These two resonances correspond to
hyperfine components of the same weakly bound singlet state. In
fact at zero field the two states are labeled by $F=0$ and $2$,
where $F$ is the vector sum of the atomic spin $f$ of the two
atoms. Similar hyperfine structure has been observed in
near-threshold singlet sodium spectroscopy~\cite{Samuelis:2001}.
Dieckmann \textit{et al.}~\cite{Dieckmann} have observed two loss
features in this magnetic field region, one broad and one narrow.
The narrow feature at 550 G is in agreement with the position of
our predicted narrow resonance, but the width is an order of
magnitude larger than we predict. No uncertainties are given for
the positions of the resonances as these are preliminary results.

In conclusion, we have observed a zero crossing in the elastic
scattering length of a mixture of the two lowest hyperfine states
of $^6$Li. The observed broad loss feature near 650 G is not
centered on the broad two-body Feshbach resonance whose zero
crossing is located at $528\pm4$ G and whose theoretical maximum
is centered near 860 G. We have used the measured zero crossing to
further constrain the singlet potential and also predict a second,
narrow Feshbach resonance at approximately 550 G. The observation
of this resonance will require magnetic field resolution beyond
the capability of our present system. Further, the cross section
near 860 G is unitarity limited at the trap depths and
temperatures employed in the present experiments. The evaporation
rate is therefore insensitive to  the scattering length, and
nearly independent of magnetic field near the resonance. Since a
Feshbach resonance does appear to exist in the magnetic field
region below 1 kG and the loss rates occur over time scales of a
second at the densities and temperatures of interest, mixtures of
the two lowest-lying hyperfine states of $^6$Li appear promising
for studies of Fermi superfluidity.

Note added: The Duke experimental and NIST theory groups wish to
acknowledge experimental input from R. Grimm, whose group has made
similar measurements of the zero crossing of the scattering length
in collaboration with A. Mosk and M. Weidm\"{u}ller. Their results
closely agree with measured data reported here.

The Duke research is supported by the Physics divisions of the
Army Research Office and the National Science Foundation, the
Fundamental Physics in Microgravity Research program of the
National Aeronautics and Space Administration, and the Chemical
Sciences, Geosciences and Biosciences Division of the Office of
Basic Energy Sciences, Office of Science, U. S. Department of
Energy. The NIST effort is supported in part by the Office of
Naval Research.


\begin{references}
\bibitem{Holland}M.~Holland, S.~Kokkelmans, M.~L.~Chiofalo,
and R.~Walser, Phys. Rev. Lett. {\bf 87}, 120406 (2001).
\bibitem{Timmermans}E. Timmermans, K. Furuya, P. W. Milonni, A.
K. Kerman,  Phys. Lett. A {\bf 285}, 228 (2001).
\bibitem{Griffin}Y. Ohashi and A. Griffin, cond-mat/0201262
(2002).
\bibitem{Kokkelmans}S. Kokkelmans, J. N. Milstein, M.
L. Chiofalo, R. Walser, and M. J. Holland, Phys. Rev. A 65, 053617
(2002).
\bibitem{JinFeshbach}T. Loftus, C. A. Regal, C. Ticknor,
J. L. Bohn, and D. S. Jin, Phys. Rev. Lett. 88, 173201 (2002).
\bibitem{Jin}B. DeMarco, and D. S. Jin, {\it Science} {\bf 285}, 1703 (1999).
\bibitem{Roati}G. Roati, F. Riboli, G. Modugno, and M. Inguscio,
cond-mat/0205015 (2002).
\bibitem{Hulet} A. G. Truscott, K. E. Strecker, W. I. McAlexander,
G. B. Patridge, and R. G. Hulet, {\it Science}, {\bf  291},
2570-2572 (2001).
\bibitem{Salomon}F.  Schreck, L. Khaykovich, K. L Corwin, G. Ferrari,
T. Bourdel, J. Cubizolles, and C. Salomon, Phys. Rev. Lett. {\bf
87}, 080403 (2001).
\bibitem{Granade}S. R. Granade, M. E. Gehm, K. M. O'Hara, and J.
E. Thomas, Phys. Rev. Lett. {\bf 88}, 120405 (2002).
\bibitem{Hadzibabic}Z. Hadzibabic, C. A. Stan, K. Dieckmann, S.
Gupta, M. W. Zwierlein, A. G\"{o}rlitz, and W. Ketterle, Phys.
Rev. Lett. {\bf 88}, 160401 (2002).
\bibitem{Verhaar}See E. Tiesinga, B. J. Verhaar, and H. T. C.
Stoof, Phys. Rev. A {\bf 47}, 4114 (1993) and references therein.
\bibitem{Elastic}M. Houbiers, H. T. C. Stoof, W. McAlexander, and R. Hulet, Phys.
Rev. A 57, R1497 (1998).
\bibitem{s-wave}E. R. I. Abraham, W. I. McAlexander, J. M. Gerton,
R. G. Hulet, R. Cote, and A. Dalgarno, Phys. Rev A {\bf 55}, R3299
(1997).
 \bibitem{Kokkelmans2}The analytic model was provided by S. Kokkelmans.
\bibitem{O'Hara1}K. M. O'Hara, S. R. Granade, M. E. Gehm, T. A. Savard,
S Bali, C. Freed, and J. E. Thomas, Phys. Rev. Lett. {\bf 82},
4204 (1999).
\bibitem{O'Hara2}K. M. O'Hara, M. E. Gehm, S. R. Granade, S. Bali,
and J. E. Thomas, Phys. Rev. Lett. {\bf 85}, 2092 (2000).
\bibitem{Dieckmann}K. Dieckmann, C. A. Stan, S. Gupta, Z.
Hadzibabic, C. H. Schunck, and W. Ketterle, cond-mat/0207046
(2002).
\bibitem{Cote:1994} R. C\^{o}t\'{e}, A. Dalgarno, and M.J. Jamieson,
Phys. Rev. A \textbf{50}, 399 (1994).
\bibitem{Linton:1999} C. Linton, F. Martin, A.J. Ross, I. Russier,
P. Crozet, A. Yiannopoulou, Li Li, and A.M. Lyyra, J. Mol.
Spectrosc. \textbf{196}, 20 (1999).
\bibitem{Yan:1996} Z.-C. Yan, J.F. Babb, A. Dalgarno, and
G.W. Drake Phys. Rev. A \textbf{58} 2824 (1996).
\bibitem{vanAbeelen:1997} F.A. van Abeelen, B.J. Verhaar, and A.J. Moerdijk,
Phys. Rev. A \textbf{55}, 4377 (1997).
\bibitem{Samuelis:2001} C. Samuelis, E. Tiesinga, T. Laue, M. Elbs,
H. Kn\"{o}ckel, and E. Tiemann, Phys. Rev. A \textbf{63}, 012710
(2001).

\end{references}
\end{document}